# A Byzantine Fault Tolerance Approach towards AI Safety

**John deVadoss**
Global Blockchain Business Council
Washington, DC
*johnd@ieee.org*

**Dr. Matthias Artzt**
Deutsche Bank
Frankfurt, Deutschland
*matthias.artzt@db.com*

## Abstract

Ensuring that an AI system behaves reliably and as intended —especially in the presence of unexpected faults or adversarial conditions— is a complex challenge. Inspired by the field of *Byzantine Fault Tolerance* (BFT) from distributed computing, we explore a fault-tolerance architecture for AI safety. BFT refers to a system's ability to continue operating correctly even when some components misbehave arbitrarily or maliciously. In classical terms, a Byzantine fault-tolerant system can withstand up to $f$ faulty nodes out of $N$ as long as $N >= 3f + 1$. By drawing an analogy between unreliable, corrupt, misbehaving or malicious AI artifacts and *Byzantine nodes* in a distributed system, we propose an architecture that leverages consensus mechanisms to enhance AI safety and reliability.

## 1  Introduction

As AI models and systems advance, concerns about their potential risks and long-term effects grow. While the industry promotes the idea of "safe" AI with built-in protections and alignment with human values and ethical principles, recent developments suggest that achieving such so-called safe AI will be a rocky road. We summarize the commonly used approaches towards AI safety, their shortfalls, and propose a radically new approach [1] for the industry.

### 1.1  State-of-the-art safety techniques and their shortcomings

Refusal behaviors in AI systems are ex ante mechanisms ostensibly designed to prevent frontier AI models, i.e., large language models (LLM), from generating responses that violate safety guidelines, ethics, or other undesired behavior. These mechanisms are typically realized using so-called guardrails that recognize certain prompts and requests, including terms and phrases, as harmful. In practice, however, prompt injections [2] and related jailbreak attacks enable bad actors to manipulate the model's responses by subtly altering or injecting specific instructions within a prompt.

In the context of LLMs, the latent space [3] is a mathematical representation capturing the underlying patterns and features of the training data. It is essentially a compressed, lower-dimensional space where different points represent various aspects of learned information. By manipulating these points, an LLM can generate diverse outputs. One strategy for enhancing AI safety involves modifying the model's parameters to constrain its latent space. However, this method typically proves effective only along one or a few specific directions within the latent space, making the model still susceptible to further parameter manipulation by malicious actors.

Formal verification of AI models involves the use of mathematical methods to prove—or attempt to prove—that an AI system will behave correctly and safely within defined parameters. However, since models like large language models (LLMs) are inherently stochastic, verification techniques for such systems tend to focus on providing probabilistic guarantees. Methods such as Monte Carlo simulations are commonly employed, but they can only offer statistical assurances rather than definitive proof. Moreover, formal verification struggles to scale effectively to large and complex systems, further limiting its practicality as frontier AI models continue to grow in sophistication.

Last, but not least, as LLMs get more and more powerful, it is very apparent that the models exhibit emergent behaviors [4], such as 'faking' alignment with the safety rules and restrictions that are imposed on their behavior. Latent behavior in such models remains an underexplored area of research—particularly when it comes to deceptive behavior, which is still not fully understood by the research community. As these models continue to scale in complexity and capability, it is reasonable to expect that the scope and impact of latent behaviors will also grow.

## 1.2   New approaches to address the safety flaws

One approach is to use the metaphor of the "Oracle"' as the preferred way to use AI models, i.e. ask questions and interpret the responses within the safety frameworks in place. The Oracle at Delphi was consulted for insight; it was not asked to act on behalf of the requester. Similarly, frontier LLMs may be utilized as founts for expert guidance, insight, advice, and understanding, without execution being delegated to the models. This approach becomes particularly relevant as the industry is now beginning to evangelize the idea of AI agents, with varying degrees of autonomy.

Another approach involves deploying a second layer of AI models to oversee and regulate the behavior of the base large language model (LLM). A simplified example of this method is the use of domain-specific, highly specialized models that monitor the base LLM, helping to detect undesired, malicious, or otherwise unsafe behaviors as defined by a given safety framework. A key challenge, however, is ensuring that this second layer remains impervious to manipulation or influence by the base LLM.

Our recommendation is that the industry use what is known as Byzantine Fault Tolerance (BFT) [5] in the field of Computer Science. Byzantine Fault Tolerance is a consensus-based, algorithmic technique to ensure that the overall system can still function correctly even if some parts fail or act maliciously. Byzantine Fault Tolerance is a consensus-based algorithmic technique that ensures a system can continue to operate correctly even if some components fail or act maliciously. Rather than focusing on aligning and securing a single large language model (LLM) in isolation, BFT can be applied across multiple peer-level AI models in mission-critical scenarios. This ensures that the collective outputs and responses of these models remain safe and reliable. If one or a subset of models produces faulty, incorrect, or malicious data, the majority consensus among the other models can override it, leading to a safer overall outcome. Furthermore, we recommend designing and training these peer-level AI models with heterogeneity in mind—incorporating diverse software and hardware strategies, including different algorithms, architectures, hardware platforms, interconnects, and training datasets—ideally sourced from a variety of vendor

## 2      System Architecture

At a high level, the proposed system architecture treats an AI system as a collection of redundant, cooperating artifacts or modules rather than a single monolithic system. Each critical decision or prediction is produced not by one component alone, but by multiple parallel components (optimally diverse in design) that collectively decide the outcome. The architecture is inspired by the classical BFT setup of N replicated nodes processing the same inputs and then running a consensus protocol to agree on the correct output. In an AI context, this could mean deploying several independent models or agents that all receive the same input (e.g. sensor data, user query) and propose an action or answer. A coordination mechanism then compares and combines their outputs to produce a single robust decision.

In practice, such an architecture might consist of *3f + 1* modules to tolerate up to *f* faulty ones (following the BFT resilience criterion). For example, to tolerate one arbitrary failure, four parallel AI modules would be used. These modules could be homogeneous (the same model replicated) or heterogeneous (different models or algorithms solving the same task) – a design choice discussed further below. The system also includes a consensus layer that orchestrates agreement among modules. This layer can be implemented either as a dedicated voter component or in a fully distributed manner where the AI modules themselves exchange messages and vote on the result. A simple approach is a majority vote on outputs, but a more robust approach involves a Byzantine consensus algorithm that can detect and exclude malicious or inconsistent contributors.

Crucially, the architecture emphasizes fault isolation and containment. Each AI module should be isolated such that failure in one does not directly corrupt others. By isolating modules and only allowing interaction through the consensus mechanism, we prevent a faulty module from directly manipulating the state of others – it can only influence the final outcome by attempting to sway the vote, which will fail if the majority (or supermajority) are correct. Additionally, to avoid common-mode failures (where a single bug or vulnerability affects all modules identically), the architecture encourages design diversity. This could involve using different model architectures, training data or algorithms for each redundant module so that they won't all make the same mistakes. Diversity ensures that even if one module encounters a corner-case leading to unsafe behavior, others are less likely to suffer the exact same flaw.

## 3      Implementation Strategies

Implementing a BFT-inspired AI safety architecture requires careful consideration of both software and hardware design. One straightforward strategy is active replication: run multiple instances of the AI artifacts in parallel on separate hardware or isolated computing environments. All instances receive identical inputs and are triggered to produce an output for each decision cycle. If all outputs agree, the decision is accepted; if there is a discrepancy, a voting or consensus process is invoked to determine the correct result. This mirrors the approach used in some aerospace systems—for example, the Space Shuttle's flight control system famously ran four redundant computers in sync [6] and would "vote off" any computer that disagreed with the others, with a fifth independent computer as a backup for added safety. Such real-world implementations validate the viability of redundancy and voting for safety-critical decisions.

### 3.1 Redundancy Management

In implementing redundancy, one must decide how to handle disagreements among modules. A simple majority vote (e.g., choosing the output that at least 2 out of 3 modules agree on) is one approach. However, in high-risk AI scenarios, outputs might not be binary or even easily comparable (consider differing explanations or control actions). One implementation strategy is to reduce outputs to a common decision space for voting – for instance, a set of high-level action commands or a classification label. Alternatively, the system might designate a leader module for each decision that others validate, akin to a primary server in Practical BFT systems. If the leader's output is validated by the others (quorum), it is adopted; if not (the leader is faulty), a new leader is chosen. This leader rotation strategy can increase efficiency by not always requiring full multi-way comparison, at the cost of adding complexity for leader election and change.

### 3.2 Diversity and N-Version Programming

As mentioned, having identical copies of a model is risky because they could all fail in the same way. An implementation strategy to introduce diversity is N-version programming, where multiple teams develop functionally equivalent modules with different designs. In the context of AI, this could mean training multiple models with different architectures (e.g., decision tree, neural network, rule-based system) for the same task, or training the same neural architecture on different subsets of data or with different initialization (as in deep ensembles). Deep ensemble techniques are known to improve reliability by combining outputs of models that make independent errors [7]. For example, one could deploy an ensemble of neural networks each with slightly varied training data and then use a weighted voting scheme on their outputs. The system should be implemented to allow easy integration of diverse modules and to collect their results for voting. A majority voter component can be implemented as a separate software module listening to all models' outputs and computing the agreed result. Introducing a majority voter or plausibility checker is essential to effectively combine multiple safety mechanisms or detectors.

### 3.3 Communication and Overhead

Another consideration is the communication overhead and synchronization of decisions. If the modules run on separate machines (e.g., a distributed robotic swarm or cloud-based ensemble), we need a network protocol to exchange their outputs or votes. Low-latency communication frameworks (such as gRPC [8]) can facilitate this. The system designer must ensure that all modules base their decision on the same input – this might involve broadcasting sensor readings or user requests to all replicas simultaneously. To keep replicas in sync, especially if the decision-making is an ongoing process (streaming data), one might implement a lock-step execution: all modules process input frame t, then pause until consensus is reached, then move to t + 1, etc. This ensures a coherent state across the redundant system. The downside is increased latency due to the synchronization and voting time; implementation strategies to mitigate this include pipelining (overlapping execution with the previous consensus) or optimistic execution (proceeding and only rolling back if consensus fails, akin to speculative execution). In safety-critical scenarios, however, deterministic execution and synchronization are usually preferred to ensure clarity on the system state at each decision point.

### 3.4    Fault Detection and Recovery

Implementation should also handle the case when a module is identified as faulty. In BFT systems, if one replica is consistently outvoted by others, it can be isolated or restarted. The AI safety system could similarly include a supervisor that monitors each module's performance and consistency with peers. If a module deviates too often from the consensus, it could be flagged for maintenance (or automatically rebooted/replaced by a fresh instance). This is analogous to fault containment units [9] in traditional fault-tolerant design. Additionally, a mechanism for state recovery is needed if a module crashes and rejoins – it must obtain the current state of the system (which the other replicas can provide through a state transfer protocol) so that it can resume participating in consensus correctly. Implementing state transfer in AI might involve sending the latest model parameters or relevant memory of the task – in many cases, if each decision is independent (e.g., classification per image frame), then a restarted module can simply start fresh on the next input. But if the AI has an evolving state (e.g., an agent with memory or an ongoing plan), consensus on state updates should be part of the protocol (just as BFT protocols replicate state machine transitions).

## 4      Consensus Algorithms for Robust Agreement

At the heart of the BFT-inspired safety approach are the *consensus algorithms* that allow multiple AI modules to agree on one output. Consensus algorithms have been studied for decades in distributed systems to ensure **consistency** even when some components are unreliable. In our AI safety context, a consensus algorithm coordinates the redundant modules such that the overall system produces a single, agreed-upon action or prediction, with guarantees that this decision has been vetted by a quorum of non-faulty modules. Here we outline how classical BFT consensus algorithms can be applied and adapted, and we compare possible algorithms.

### 4.1    Byzantine Agreement Basics

A Byzantine agreement protocol must satisfy two key properties: **safety** (no two non-faulty modules decide on different outputs) and **liveness** (the modules eventually reach a decision if a majority are functioning). The algorithms typically require multiple rounds of message exchange. A fundamental result in BFT theory is that reaching consensus with *f* Byzantine failures requires at least *3f + 1* participants and a quorum of at least *2f + 1* to agree. In other words, the correct modules must outnumber faulty ones by a sufficient margin to both have a majority and to intersect any two quorums (so that one honest member is common to all quorums). For example, with 1 tolerated fault (f=1, N=4), a decision is considered committed when at least 3 modules (2f + 1 = 3) agree on it, ensuring any other set of 3 (for a different outcome) would share at least one honest module and thus could not all be valid simultaneously.

## 4.2  Practical Byzantine Fault Tolerance (PBFT)

One of the best-known BFT consensus algorithms is PBFT [5] by Castro and Liskov. PBFT is a *leader-based* protocol that operates in **three phases: Pre-Prepare, Prepare, Commit**. Adapting this to AI modules: one module (leader) proposes an output (this is analogous to a client request or transaction in distributed systems). In the *Pre-Prepare* phase, the leader broadcasts its proposed output to all other modules. In *Prepare*, all modules (including the leader and backups) exchange messages indicating they received the proposal and are prepared to accept it; they build a certificate (proof) that a certain number of peers (at least 2f + 1) have seen the same proposal. In *Commit*, the modules broadcast a commit message once they have enough prepare confirmations, ensuring that a sufficient quorum has agreed, and then they *commit* to the proposed output. After the commit phase, all non-faulty modules will have agreed on the same output and can act on it (e.g., execute the decided action). The client (or the decision module) can likewise wait for *f + 1* matching responses from different modules to be confident of the decision (since at least one of those must be from a non-faulty module, implying the decision is the agreed one). This three-phase design guarantees that even if the leader was malicious (proposing conflicting things to different modules, etc.) or some backups are lying, the correct modules will only commit to one consistent outcome. If the leader fails to get agreement (perhaps because it was faulty), the protocol has a **view change** sub-protocol to select a new leader and retry the consensus, ensuring liveness.

For the AI safety system, a full PBFT implementation imposes additional constraints but provides stronger guarantees. In scenarios where performance is critical and the system is small (say 4 or 7 modules), the overhead of three rounds of communication may be acceptable (modern optimizations and local networks can make PBFT quite fast). An alternative is to use simpler consensus when possible. For instance, if outputs are easily comparable and quick to verify, a simplified *Byzantine quorum vote* might suffice: each module broadcasts its output's hash, and if all hashes match, it's done; if not, they send their full outputs and each module picks the majority value (this is a simpler two-round agreement). However, simpler voting schemes lack some guarantees: a malicious module could try to equivocate (tell different modules different results), but a robust BFT protocol with signed messages would catch this inconsistency. Thus, for high assurance, implementing a standard BFT algorithm (like PBFT or one of its many modern derivatives) is advisable.

## 4.3  Consensus Variants and Considerations

It is worth noting that in distributed systems literature, many consensus algorithms exist. *Crash-fault tolerant* algorithms like Paxos [10] and Raft [11] assume nodes only fail by stopping, not by malicious acts. Those are insufficient for our needs since an AI module might produce actively misleading output (equivalent to a malicious fault). Therefore, Byzantine-fault tolerant algorithms (PBFT [5]) are the relevant ones. Some blockchain consensus mechanisms (e.g., Hyperledger's consensus [12]) are essentially BFT protocols optimized for certain network conditions and could potentially be adapted if the AI modules are distributed over a network. In real-time AI systems, a partially synchronous timing model (assuming messages are delivered reasonably fast most of the time) is usually acceptable, which is the model PBFT works under. If timing is critical, one might impose timeouts: e.g., if no consensus by a deadline, assume leader failed and switch.

One important aspect to implement is the **quorum validation** of the final decision. To illustrate, suppose we have 5 AI modules and tolerate f=1 faulty. The consensus mechanism should ensure that at least 4 modules have agreed on the output (since $2f + 1 = 3$, but to be safe under all races, often systems require all non-faulty to commit). The output is then considered final. The system can be designed such that the *action* (e.g., a control command to a robot) is only executed after at least 3 (or 4) modules sign off on it. This way, even if one module is compromised and tries to issue a dangerous command, that command will not reach execution unless others independently agree with it.

In summary, consensus algorithms provide the formal underpinning that makes the multi-module architecture effective: they guarantee that the non-faulty parts of the AI reach the same decision and that any faulty part cannot manipulate the outcome. This significantly enhances safety, as *agreement on a correct outcome* is assured as long as the majority of modules are correct.

# 5   Use Case Examples

To ground the discussion, we consider several use cases where a Byzantine fault-tolerant approach can enhance AI safety:

## 5.1   Autonomous Vehicle Decision-Making

Self-driving cars rely on AI for perception and planning, but sensor failures or software bugs can lead to catastrophic errors. A BFT-inspired safety system in an autonomous vehicle might run multiple independent perception modules (e.g., vision, LIDAR, radar processing units) and multiple planning algorithms in parallel. These modules would all evaluate the current driving situation and propose an action (e.g. "brake", "continue", "swerve left"). Through a consensus mechanism, the vehicle's central controller accepts an action only if a quorum of modules agree it is safe. For instance, if 4 out of 5 decision modules concur that an object on the road is a harmless plastic bag and it's safe to continue, but 1 module misinterprets it as a person and votes to brake, the consensus outcome will be to continue driving (since the one dissenting module is outvoted by the majority). Conversely, if one module fails to detect an obstacle but others do see it and vote to stop, the car will stop. This redundancy greatly reduces the chance of single-point failures causing accidents. Notably, *design diversity* can be applied: the 5 modules might use different sensor data or algorithms (one focusing on vision cameras, another on radar) so that no single fault (like a blinded camera) knocks out all detection.

This approach parallels aviation systems where multiple avionics computers cross-check each other for flight control, providing a proven template for reliable autonomy.

## 5.2   AI Assistants

Large language model-based assistants sometimes produce inappropriate or unsafe responses. To enhance reliability, one could deploy *multiple AI agents in parallel for content generation*, each with slightly different training or prompting, and a consensus checker. Suppose a user asks a complex question. Three separate instances of an AI model generate answers. Before any answer is shown, a *consensus-checking step* is applied: the answers are compared for consistency and adherence to safety guidelines. If one answer contains disallowed or dubious content that the others avoided, the system can flag that instance as potentially faulty and rely on the "safer" majority answers. The final response could even be an aggregate or vetted combination of the outputs. This process ensures that no single instance's lapse (such as using a forbidden phrase or revealing confidential information or personal data to be treated as sensitive from the data protection point of view) makes it to the user, since the other agents serve as a check. In essence, the AI agents *verify each other's responses for accuracy and ethical compliance before finalizing an answer*.

This use case illustrates how consensus can enforce not just correctness but also alignment with human values by treating a deviation as a Byzantine fault.

### 5.3 Multi-Agent Robotics and Swarms

In scenarios with multiple AI-driven agents (drones, robots), Byzantine fault tolerance can safeguard coordination. For example, a drone swarm can use a consensus algorithm to agree on a formation or on which target to track, so that even if a few drones are malfunctioning or malicious, the majority's decision prevails and the swarm does not fracture. Similarly, a team of AI-powered vehicles at an intersection could collectively decide which one goes first to avoid collisions—if one vehicle's AI tries to cheat or is faulty, the others will override that through consensus. While this ventures into decentralized multi-agent territory, the principles are the same: redundant agents negotiating a common outcome. In fact, research in multi-agent systems has started to incorporate blockchain and BFT consensus to ensure agreement on shared tasks or world state [13]. This use case highlights that BFT approaches can scale out to networked AI systems, not just components within one system.

Each of these examples demonstrates a pattern: **redundancy + agreement = safety**. By having multiple artifacts) weigh in, the system is less likely to converge on a faulty prediction or outcome. And by requiring consensus, the system gains assurance that the action it takes is corroborated by independent analyses. This significantly reduces the risk of an unsafe outcome due to any single point of failure.

# 6     Trade-offs and Comparisons with Alternative Approaches

While a Byzantine fault tolerance-inspired approach can greatly enhance the reliability and safety of AI systems, it is not without trade-offs and alternatives. It's important to critically examine the costs of this approach and how it compares to other AI safety techniques.

### 6.1    Performance vs. Safety Overhead

One clear trade-off is computational and latency overhead. Running $3f + 1$ copies of an AI module to tolerate $f$ faults means a substantial increase in required resources. For instance, tolerating even 1 fault demands 4x the computation in the worst case. This may be acceptable for small models or when safety is paramount (as in spacecraft or medical devices), but it could be impractical for extremely large models or real-time constraints. The consensus process itself also introduces latency – decisions are not immediate but require collecting votes or multiple rounds of communication. In scenarios like autonomous driving, where milliseconds matter, designers must ensure the consensus protocol is optimized (e.g., by using parallel hardware and fast message passing). This is a classic *latency vs. safety* trade-off: the system might react slightly slower, but with higher confidence in correctness. Techniques like pipelining consensus, as mentioned, or using a *2-out-of-3 voter* instead of a full 3-phase protocol for very quick decisions, can mitigate latency at some potential cost to fault tolerance strength.

## 6.2 Complexity and Engineering Effort

Another trade-off is the complexity of implementation. Integrating a BFT consensus layer into an AI system is non-trivial. It adds complexity in system architecture, which itself can introduce new risks (e.g., bugs in the voting logic, or synchronization issues). By contrast, simpler safety measures (like a single software watchdog or a heuristic monitor that checks if outputs are within reasonable bounds) might be easier to implement, though less reliable. The BFT approach also requires careful *synchronization of replicas* and handling of edge cases like what happens if modules disagree – these are essentially new failure modes that need testing. In comparison, an alternative approach such as *formal verification* of an AI decision module would attempt to mathematically prove that the module cannot enter an unsafe state. For complex learning systems, formal verification is often impractical, which is why the redundancy approach is considered.

## 6.3 Consensus Robustness vs. Single-Model Robustness

Traditional AI safety research often focuses on making a *single model* robust through methods like adversarial training, robust optimization, or uncertainty estimation. The BFT-inspired approach, on the other hand, assumes any single model *can* fail and instead focuses on system-level robustness. A comparison can be drawn here: improving a single model's robustness is like strengthening a single engine in a rocket, whereas BFT is like having multiple engines so that even if one fails, the rocket still flies. Ideally, one could **combine** these approaches: use adversarial training and safety-focused design for each module to reduce its failure probability and have redundancy so that any failures that still occur are caught by others. The trade-off is that developing and maintaining multiple diverse models with robust training is a heavy burden on the AI development pipeline.

## 6.4 Common-Mode Failures and Diversity

A subtle challenge in the BFT approach is ensuring that modules fail independently. If all modules are identical and an unexpected condition occurs (e.g., a novel obstacle type on the road that confuses the model), there is a risk that *all replicas make the same mistake*, thus defeating the purpose of redundancy. Byzantine fault tolerance theory assumes faults are independent up to $f$ malicious actors; if a single fault (like a design flaw) can simultaneously flip more than $f$ modules, the guarantee is void. This is why **design diversity** is often touted as an essential companion to replication. However, achieving true diversity has trade-offs: different model designs may have different performance characteristics, and combining their outputs might be non-trivial. Additionally, diverse implementations mean a larger codebase and more complex validation. In practice, engineers must balance between using sufficiently distinct methods and maintaining manageability. The Space Shuttle example again is instructive: the primary four computers all ran the same software (for exact sync), which meant a design bug could affect all four; they mitigated this by a fifth computer with independently written software as a *diverse backup*. In AI terms, one might deploy one algorithm that's completely different (say a simpler rule-based system) alongside neural networks, to handle situations the networks might all misjudge. This is effectively a hybrid approach combining *fault tolerance* with *fault avoidance* (by not putting all eggs in one algorithmic basket).

## 6.5 Accuracy vs. Safety Voting Thresholds

Using consensus introduces interesting trade-offs in terms of sensitivity. For example, if we require a strict majority (e.g., 3 of 5 modules) to flag a condition as hazardous, it might reduce false alarms but increase the chance of missing a real issue that only one highly specialized module detects. Conversely, if we acted on a single module's alarm (1 out of 5), we maximize sensitivity but lose fault tolerance (one faulty alarm could trigger an unnecessary action). A compromise is requiring *2 out of 3* or *N-1 out of N* agreements for certain critical signals. In fact, safety engineers often design voters like "two-out-of-three" systems to balance false positives and false negatives. For instance, a 2oo3 (two-out-of-three) voter will only trigger if at least two modules agree on the condition. This reduces spurious triggers (since it's unlikely two fail in the same way at the same time spuriously) and still catches most real issues (as long as at least two modules can detect it). However, it could miss subtle issues that only one module picks up (higher false negative risk). Setting these thresholds is a trade-off decision: do we prioritize avoiding **false positives** (unnecessary interruptions, which might degrade user trust or efficiency) or avoiding **false negatives** (missed detections of a problem)? BFT consensus protocols typically opt for avoiding false negatives in the sense that they won't decide on an action unless a quorum agrees – aligning with a conservative approach to safety (don't do it unless reasonably sure). Designers can tune the consensus (for example, require unanimity for extremely sensitive actions, majority for routine ones) as part of the safety strategy.

## 6.6 Comparison with Other Safety Approaches

Aside from robust training and formal methods, there are other approaches like *monitoring and sandboxing* [14], and *human-in-the-loop oversight* [15]. A monitoring system might watch an AI's inputs and outputs for anomalies (out-of-distribution inputs, or actions that violate predefined constraints). This is a single-checker approach rather than multiple agents reaching consensus. Its advantage is simplicity – one well-crafted monitor can catch many issues. However, the monitor itself can be seen as a single point of failure (if it misses something or fails, the system has no backup). Our BFT approach effectively distributes the monitoring across peers. Human oversight (having a human approve or veto AI decisions) is another safety layer, but it does not scale to systems that need split-second decisions and can introduce its own delays and inconsistencies. In scenarios where human oversight is feasible (like an AI medical diagnosis tool where a doctor reviews results), one might not need full BFT redundancy; instead, the human serves as the ultimate check. But in fully autonomous systems, human oversight at run-time is unavailable, so machine redundancy is the fallback.

## 6.7 Legal aspects when utilizing BFT

It is pivotal to diligently vet the legal implications of BFT usage. Those may arise from various angles: one is the liability risk when setting up multiple AI modules. In the case all modules are contaminated as when they are trained with contaminated data, the liability exposure may be amplified depending on the number of modules being used. This issue crystallizes when training or prompting the AI modules with data of individuals. The key backbone of all advanced privacy regimes, such as the GDPR [16] in Europe, is the data minimization principle which stipulates to only use the data which is required to achieve a given purpose and delete the data when it is not needed any more. It is apparent that this principle will clash with the BFT approach if all AI modules are being fed with the same set of personal data simultaneously. Another aspect which is worthwhile taking into consideration is the right to be forgotten which is, by nature, not compliant with AI systems. To that end, the only way to mitigate the risk of being privacy in-compliant is to anonymize the personal data before submitting it to the array of AI modules since anonymization renders that sort of data non-personal.

It's also worth noting that BFT addresses **internal faults and consensus**, but it does not inherently guarantee optimal performance or correct goals. If all modules are systematically biased or if the majority shares a flawed objective, they could *consistently agree on a bad decision*. Thus, BFT should complement, not replace, efforts in ensuring the AI artifacts have the correct objective or reward function. For example, if all redundant modules are trained with a reward function that undervalues safety in rare scenarios, they might all converge on an unsafe action that is nonetheless consensual. So, while BFT can mitigate random and adversarial errors, it should be combined with proper alignment techniques and testing of the AI's decision criteria.

In summary, the BFT-inspired safety approach trades higher resource usage and complexity for a significant gain in resilience to faults. Compared to alternatives like single-model robustness improvements, it provides a more *systemic guard* against failures, but it does not eliminate the need for good design of each component. The best strategy might integrate multiple approaches: use BFT consensus for critical decisions while also training each model to be as robust as possible and employing simpler safety checks as additional layers (defense in depth). This layered approach is often recommended in safety engineering: multiple redundant mechanisms, some aiming to prevent faults and others to tolerate them. Our focus here has been on the tolerance layer; in practice it should work hand in hand with preventive measures.

# 7      Conclusion and Future Directions

The Byzantine fault tolerance-inspired approach offers a promising pathway to enhance AI safety through architectural rigor and consensus. By structuring AI systems as ensembles of AI artifacts or modules that *check and balance* each other, we gain strong assurances that no single errant or deceptive component can easily steer the system into an unsafe state. We have detailed how such a system can be built – from the overall architecture of replicated modules and voters to the implementation strategies ensuring diversity and synchronization, to the consensus algorithms that guarantee agreement among modules. We have also discussed use cases demonstrating concrete benefits in domains like autonomous vehicles, AI assistants, and robotics, and examined the trade-offs relative to other safety techniques. The academic and practical evidence suggests that complex AI systems, much like complex distributed systems, can greatly benefit from redundancy and Byzantine fault tolerance when outright prevention of all faults is untenable.

Looking forward, several recommendations and research directions emerge:

- **Scalable Consensus for AI:** As AI systems grow in complexity, research is needed on consensus protocols that can handle *larger numbers of modules or very fast decision loops*. Classical PBFT works well up to perhaps tens of nodes; beyond that, the communication overhead grows. Future work could explore more efficient BFT algorithms (e.g., leveraging threshold cryptography or sharding of tasks among modules) that maintain safety without incurring prohibitive delays. There is also room to tailor consensus algorithms specifically for AI workloads – for example, consensus that can handle *continuous streams* of data and decisions, rather than discrete one-shot agreements, would be valuable for robotics and vision systems.

- **Automated Diversity Generation:** Ensuring diversity among redundant AI modules currently may require manual effort (training different models, writing different algorithms, creating diverse synthetic datasets). A forward-looking idea is to automate the generation of diverse models. Techniques like *neural architecture search* or *ensemble breeding* could produce a set of models that are as accurate as possible while having uncorrelated errors. Developing systematic ways to create N-version AI modules with guaranteed divergence in failure modes is an open research question. It intersects with topics of robust machine learning and could benefit from evolutionary algorithms that aim for a diverse set of solutions rather than a single optimum.

- **Integration of Learning and Consensus:** In current designs, the consensus layer is separate from the learning modules (which act as black-box decision makers). An intriguing direction is to integrate consensus *within* the learning process. For example, training multiple agents that not only make predictions but also learn to listen to their peers and come to consensus could be a way to build fault tolerance inherently. There have been proposals of using *reinforcement learning to optimize consensus protocols*, or training agents to recognize when they disagree and defer to majority. Additionally, consensus algorithms might be improved by incorporating confidence measures from AI modules – e.g., weighting votes by the self-reported uncertainty of each model (a model highly unsure of its output might effectively abstain, while confident models carry more weight). Such *weighted Byzantine consensus* could potentially reduce errors of majority when one module clearly has a better view of a situation (for instance, a radar might be more reliable than a camera in heavy fog, so the system should weight their outputs accordingly). Research into dynamic trust or weighting mechanisms in consensus (sometimes called *Byzantine fault-tolerant sensor fusion*) is a promising avenue.

- **Human and Ethical Oversight Layers:** Another recommendation is to blend the BFT-inspired technical solution with human oversight and ethical reasoning components for a comprehensive safety system. For example, one can imagine a future AI governance system where multiple AI agents and possibly a human overseer form a *joint consensus* on critical decisions (this is analogous to having a human in a multi-agent "meeting" where AI agents propose options). In safety-critical operations, an AI consensus might trigger a human review if the decision is high-risk and time permits. Establishing formal methods for *escalation* (when the AI consensus cannot be reached or detects anomalies, it hands off to a human or a safe-mode policy) will be important. This requires not only technical work but also policy and interface design so that humans can effectively understand and intervene when needed.

In conclusion, the Byzantine fault tolerance-inspired approach is not a silver bullet, but it is a powerful *structural safeguard* that can significantly enhance the trustworthiness and safety of AI systems. By accepting that components may fail, and further that frontier AI models may deceive, and designing the system to tolerate such failures, we embrace a time-honored engineering principle for a new era of intelligent machines. The approach synergizes with, rather than replaces, other safety measures: it provides a resilient backbone on which safer AI behavior can emerge. We encourage further exploration and interdisciplinary work to refine this approach, reduce its costs, and apply it to the next generation of AI applications that society will depend on. With careful design, testing, and continued innovation, Byzantine fault tolerance can become a cornerstone of AI safety in practice, much as it has been for decades in mission-critical distributed systems.